\documentclass{aa}
\usepackage{graphicx}
\usepackage{txfonts}
\usepackage{pdfpages}
\begin{document} 

    \title{Small-scale dynamo in cool main sequence stars.}

   \subtitle{II. The effect of metallicity}

   \author{V.~Witzke\inst{1}\fnmsep\thanks{e-mail: witzke@mps.mpg.de}
          \and
          H.~B.~Duehnen\inst{1}  \and 
          A.~I.~Shapiro\inst{1} \and
          D.~Przybylski\inst{1}\and
          T.~S.~Bhatia\inst{1}\and
          R.~Cameron\inst{1}\and
          S.~K.~Solanki\inst{1}
          }

   \institute{Max Planck Institute for Solar System Research, Justus-von-Liebig-Weg 3, 37077 G\"ottingen, Germany\\ }

   \date{---}

  \abstract
   {All cool main sequence stars including our Sun are thought to have magnetic fields. Observations of the Sun revealed that even in quiet regions small-scale turbulent magnetic fields are present. Simulations further showed that such magnetic fields affect the subsurface and photospheric structure, and thus the radiative transfer and emergent flux. Since small-scale turbulent magnetic fields on other stars cannot be directly observed, it is imperative to study their effects on the near surface layers numerically. }
   {Until recently comprehensive three-dimensional simulations capturing the effect of small-scale turbulent magnetic fields  only exists for the solar case. A series of investigations extending SSD simulations for other stars has been started. Here we aim to examine small-scale turbulent magnetic fields in  stars of solar effective temperature but different metallicity.}
   {We investigate the properties of three-dimensional simulations of the magneto-convection in boxes covering the upper convection zone  and photosphere carried out with the MURaM code for  metallicity values of $ \rm M/H = \{-1.0, 0.0, 0.5\}$ with and without a small-scale-dynamo. }
   {We find that small-scale turbulent magnetic fields enhanced by a small-scale turbulent dynamo noticeably affect the subsurface dynamics  and significantly change the flow velocities in the photosphere. Moreover, significantly stronger magnetic field strengths are present in the convection zone for low metallicity. Whereas, at the optical surface the averaged vertical magnetic field ranges from 64G for M/H = 0.5 to 85G for M/H = -1.0.}
   {}


\maketitle
%

\section{Introduction}

Atmospheres of cool stars, that are stars with convective envelopes like the Sun, are filled with magnetic fields. Some of these fields emerge from below the stellar surface and affect the structure of the stellar atmosphere, changing the radiative emission from the star. The various manifestations of such atmospheric changes, e.g. spectroscopic and brightness variations, Ca~II and X-ray emission and their variations \citep[see, e.g.,][and references therein]{Schrijver_Zwaan2000}, are usually referred to as stellar magnetic activity.

The interest in stellar activity is by far not limited to solar and stellar physics.  Stellar brightness variability is also a limiting factor for the detection and characterisation of exoplanets by transit-photometry missions \citep{Aigrain2004,Johnson2021}. The magnetic jitter in radial velocity affects the spectroscopic detection of planets \citep{RV_state,MeunierandLagrange2019}, while the magnetic jitter in position of the stellar brightness centre impedes the astrometric detection of planets \citep{Meunier2020,Sowmya2021,Kaplan-Lipkin2021}. Recent studies also showed that magnetic features on stellar surfaces can substantially interfere with identifying the chemical composition of exoplanetary atmospheres by transmission spectroscopy \citep{Rackham2018,Rackham2019}.


The star whose surface magnetic field has been most extensively studied is the Sun \citep[see, e.g., review by][and references therein]{Sami2006}. Observations show that the most salient manifestations of solar surface magnetism such as spots, faculae, and magnetic network are modulated by the activity cycle produced by the action of the global dynamo in the interior of the Sun  \citep{Charbonneau2020}. Surprisingly, it was also revealed that even the solar regions free from any apparent manifestation of magnetic activity (i.e.~the quiet Sun) are interspersed with small-scale  magnetic fields of mixed polarity \citep[often also referred as ``internetwork'' magnetic fields, see][]{Livingston_Harvey75}, which appear to be largely uncorrelated with the solar cycle \citep[see reviews by][]{Buehler_Lagg_2013, Lites_Centeno_2014, Borrero2017, QS_LR}.

These fields lead to considerable small-scale turbulent magnetic flux to always be present at the solar surface \citep{Trujillo2004Natur}. Simulations can successfully explain their existence by the action of a small-scale turbulent dynamo (SSD), which describes the amplification of the magnetic flux by the near-surface turbulent convection \citep[see][and references therein]{Rempel_2018ApJ}. Interestingly, it was recently shown that small-scale internetwork magnetic fields amplified by a turbulent dynamo have a significant effect on the radiation emanating from the Sun \citep{Rempel2020, Yeo2020}. This raises a compelling question of whether small-scale fields can also affect photometric and spectral stellar data. There is presently no answer to this question since  nothing is currently known about the properties of the small-scale turbulent magnetic fields  on stars others than the Sun. In particular, their effect on stellar atmospheres was not considered in the widely used 1D \citep[e.g.,][]{CastelliandKurucz2003, Gustafsson2008, Husser2013} and 3D \citep[e.g.,][]{ludwig2009cfist, Beeck2013A&Afirst, Magic_2013A&A} grids of stellar atmospheric models. 

This is the second study in  a series of papers where we  overcome  this omission and investigate the properties and effects of magnetic fields generated by small-scale turbulent dynamos in stars with outer convection zones.  In a parallel study we investigate the SSD as a function of the stellar effective temperature  \citep{Bhatia_2022}, where the we focused on stars of spectral types F3V, G2V, K0V and M0V with solar metallicity. In this paper we utilised the MURaM code \citep{Vogler_Sch_2005} to simulate the action of SSD and small-scale fields in metal-rich (M/H=0.5) and metal-poor (M/H=-1.0) stars with solar temperature and surface gravity. We further investigated the backreaction of the SSD on the near-surface convection and structure of stellar atmosphere. 
For comparison we also perform simulations for the Sun (M/H=0.0). In the forthcoming papers we will consider the effects of SSD on stellar measurements.

The paper is structured as follows. The numerical model is described in Section~\ref{Sec:model}, where we give the governing equations and the simulations setup. Subsequently, we present our results in Section~\ref{Sec:results} and discuss their implications in Section~\ref{Sec:discussion}.


\section{Model}
\label{Sec:model}
In this study we follow  the ``box-in-a-star'' approach. For that, we consider a Cartesian box filled with plasma that represents a small part of a star around the optical surface, such that  the upper convection zone and the photosphere are included. 


\subsection{Governing equations and implementation}

To solve  for the dynamics and energy transport in such a region we employ the radiative 3D MHD code {MURaM} \citep{Vogler_Sch_2005,rempel_2014,rempel_2016}, which uses a fourth-order accurate conservative, centred finite scheme for the discretisation.  
For the radiative transfer a multi-group scheme with short characteristics is implemented \citep{nordlund_1982}. A pre-tabulated equation of state is used, which we generated by the  FreeEOS code \citep{Irwin_freeeos_2012}.

MURaM solves the conservative  magnetohydrodynamic (MHD) equations for a compressible, partially ionised plasma in the following form:
\begin{eqnarray}
\label{eq:gov01}
\frac{\partial \rho}{\partial t} & = & - \mathbf{\nabla}\mathbf{\cdot}\left(\rho \mathbf{u} \right) \, \\
\label{eq:gov02}
\frac{\partial(\rho \mathbf{u})}{\partial t} & = & 
-\nabla\cdot(\rho\mathbf{u}\mathbf{u})-\nabla p +\rho\mathbf{g}  + \mathbf{F}_{\rm{L}} + \mathbf{F}_{\rm{SR}}\, \\
\label{eq:gov03}
\frac{ \partial (\rm E_{HD})}{\partial t} & = &-\nabla\cdot(\mathbf{u}(\rm E_{HD}+p))+ \rho \mathbf{u} \cdot \mathbf{g}  +\mathbf{u} \cdot \mathbf{F}_{\rm{SR}} + \mathbf{u} \cdot \mathbf{F}_{\rm{L}} \\
&  & +Q_{\rm{rad}} +Q_{\rm{res}}\, \\
\label{eq:gov04}
\frac{\partial \mathbf{B}}{\partial t} & = & \nabla\times(\mathbf{u} \times \mathbf{B})
\label{eq:governing}
\end{eqnarray}
Here, $\rho$ is the density, $p$ the pressure, $\mathbf{u}$ the velocity, $\mathbf{B}$ the magnetic field, $Q_{\rm rad}$ and $Q_{\rm{res}}$ denotes the radiative heating term and the resistive heating term, and $\mathbf{g}$ the gravitational acceleration, which in this study is set to $\rm log g = 4.4378$. The force term $\mathbf{F}_{\rm{L}}$ is the Lorentz force and $\mathbf{F}_{\rm{SR}}$ denotes the semi-relativistic (Boris) correction \citep{boris1970, gombosi2002}, which is negligible for our setup. The hydrodynamic energy, $\rm E_{HD}$, is the sum of the  internal energy ($\rm E_{int}$) and the kinetic energy ($\rm E_{HD} = E_{int} + 1/2 \rho \mathbf{u}^2$).
%
%
In order to achieve numerical stability it is necessary to implement an additional diffusion scheme. The implementation in MURaM is based on a slope-limited diffusion scheme,
where the diffusive terms are implemented in terms of  fluxes of the conserved quantities across the grid-cell boundaries \cite[for a detailed description see][]{rempel_2014, rempel_2016}.


\subsection{Boundary conditions}
The simulation domain is periodic in both horizontal ($x$ and $y$) directions and has boundaries at the top and bottom ($z_{\rm top}$, $z_{\rm bottom}$). In addition to the cells within the domain, two ghost cells at the top and bottom boundaries are introduced. These are needed for the implementation of the vertical derivatives and the boundary conditions. For most variables a closed or open boundary is  achieved  using symmetric and anti-symmetric neighbouring ghost cells at the boundary. For any variable, $v$, in a symmetric implementation the two ghost cells, $\tilde{v}_{0}, \tilde{v}_{1}$ (with a tilde denoting the ghost cells),  where  $\tilde{v}_{0}$ is the one next to the boundary, are set to corresponding values in the domain $\tilde{v}_{0} = v_0, \tilde{v}_{1} = v_1$. Whereas, for an anti-symmetric case the condition reads $\tilde{v}_{0} = -v_0, \tilde{v}_{1} = -v_1$.

The top boundary permits outflows (symmetric), but inflows (anti-symmetric) are not allowed. In addition, $z_{\rm top}$ is open to vertical magnetic fields. The formulation of the bottom boundary condition affects the magnetic field generation. 
In the past, different conditions ranging from boundary conditions that ensure a self-contained dynamo problem to various settings to mimic the deep convection zone were tested \citep{rempel_2014}. Based on equipartition arguments it was  shown that some boundary conditions mimic the structure of magnetic field deeper in the convection zone more consistently \citep{rempel_2014,hotta_ssd}. Consequently, we  consider boundary conditions similar to that in \citet{rempel_2014}, where $z_{\rm bottom}$ is open to mass flux ($\rho u_{z}$)  and magnetic fields. 
We note that an open boundary condition for magnetic fields might lead to violation of the divergence-free constraint. To generally counter the build up of deviations from $\nabla \cdot \mathbf{B} = 0$, MURaM uses a hyperbolic divergence cleaning approach \citep{div_bc_clean}.

\begin{table}
\setlength\tabcolsep{5pt}
\renewcommand{\arraystretch}{1.5}
\caption{List of top and bottom boundary conditions. We note that $\rm E_{HD}$ is obtained from $E_{int} + 1/2 \rho \mathbf{u}^2$ once all quantities are set at the boundary.}              
\label{table:00}      
\centering                                     
\begin{tabular}{|c | c| c|  }         
\hline\hline                       
Quantity      &  top  & bottom    \\   
\hline\hline
density  & symmetric &  inferred from \\
         &           & entropy \& pressure \\
mass flux vertical in    &  anti-symmetric   &  symmetric  \\
mass flux  vertical out    & symmetric    &  symmetric   \\
mass flux  horizontal  & symmetric & symmetric \\ 

$\rm E_{int}$ &  symmetric     &  inferred from    \\
                 &       & entropy \& pressure \\

vertical $\mathbf{B}$ & symmetric &  symmetric \\
horizontal $\mathbf{B}$ & anti-symmetric  &  symmetric\\
\hline \hline
pressure  & --- & see eq.~\ref{eq:pghost01}-\ref{eq:pghost02} \\
entropy in & --- & prescribed value $s_{bc}$ \\
entropy out & --- & inferred from \\
            &     & density \& $\rm E_{int}$\\
 \hline\hline  
\end{tabular}
\end{table}

To maintain an on average constant effective temperature of the simulation, the entropy inflow is specified at the bottom boundary, $s_{bc}$, while the entropy down-flows are symmetric. The pressure at the boundary and at the ghost cells is treated as described for the `HD2' condition in \citet{rempel_2016}, where the gas pressure is split into a mean pressure plus fluctuations,  $p = \overline{p} + \delta p$. However, we use a slightly different damping factor $C_{\rm dmp} = 0.8$ and add additional damping proportional to magnetic pressure. Then, our extrapolation to ghost cells reads
\begin{eqnarray}
\label{eq:pghost01}
\tilde{p}_0 & =   \overline{p}_0 \cdot \frac{p_{bc}}{\sqrt{\overline{p}_0 \overline{p}_1}}  + & C_{\rm dmp} \cdot  \delta \tilde{p}_0  - 0.5 \tilde{B}_{z, 0}^2 (1-C_{\rm dmp}),\, \\
\label{eq:pghost02}
\tilde{p}_1 & =  \overline{p}_0 \cdot \frac{p_{bc}^2}{{\overline{p}_0 \overline{p}_1}}  + & C_{\rm dmp}^2 \cdot \delta \tilde{p}_0  -  0.5 \tilde{B}_{z, 0}^2 (1-C_{\rm dmp}^2), \,
\label{eq:pressure_ghost}
\end{eqnarray}
where $p_{bc}$ is the fixed pressure value that we aim to maintain at the bottom and $\tilde{B}_{z, 0}$ is the vertical magnetic field at the first ghost cell (i.e. due to the symmetric condition $\tilde{B}_{z, 0} = {B}_{z, 0}$). A list of the boundary conditions is given in table~\ref{table:00}.


\subsection{Simulation setup}
\label{subsec:sim_setup}

The horizontal extent of our simulation box is 9 Mm $\times$ 9 Mm and it is chosen to contain at least a dozen granulation cells.  The vertical extent is 5 Mm so that  the box covers the photosphere (roughly up to the temperature minimum) and reaches down in to the upper convection zone (ca. 4 Mm below the optical surface, $\tau_{500} = 1$ ). While the box has an extent of 1 Mm above the surface the top layers are affected  by the top boundary condition which is open to vertical magnetic fields. We estimate that the dynamics in roughly 0.5 Mm from the top might be influenced by this choice of boundary condition, and we make this by a grey shaded area in all plots. 
The numerical grid has 500 points in the vertical and $512 \times 512$ points in horizontal directions.
The element composition of the plasma is taken from \citet{Asplund_2009}. In this study we restrict the analysis to three different metallicity values (M/H = -1.0, 0.0, and 0.5, where compared to the Sun M/H = -1.0 corresponds to 10 times less metals and M/H = 0.5 to approximately 3 times more). 

We use a 1D mean atmospheric structures generated by the Modules for Experiments in Stellar Astrophysics (MESA) code \citep{Paxton_2011} to initialise all of the three metallicity runs. After a simulation is initialised, it is run until it reaches a saturated state. For the SSD run, we inserted a small,  random,  and flux free  magnetic field. Subsequently, the calculation was run until the root-mean-square (rms) magnetic field strength of the entire cube saturates. At the early stage, during the relaxation of the simulation,  we use grey radiative transfer. Finally, we switch to a four bin non-grey opacity table  \citep[for a more detailed discussion see][]{Voegler_2004},  where we updated the pre-tabulated opacity table according to the mean time averaged atmospheric structure several times.  

Subsequently, the entropy and pressure values at the bottom boundary are adjusted iteratively. The pressure at the bottom is chosen such that the mean optical surface is roughly four Mm above the bottom of the box. Furthermore,  the entropy inflow at the bottom is adjusted such that a time averaged mean effective temperature in the non-grey SSD run is  maintained around  5777K $\pm 15 \rm K$. The pressure and entropy conditions at the bottom, found to satisfy the above condition in the SSD run,    are kept for the pure hydrodynamic run  of the same metallicity value. 
Finally, we let the simulation evolve for 10 hours of solar time.


\begin{figure}
  \centering 
   {\includegraphics[width=0.95\linewidth]{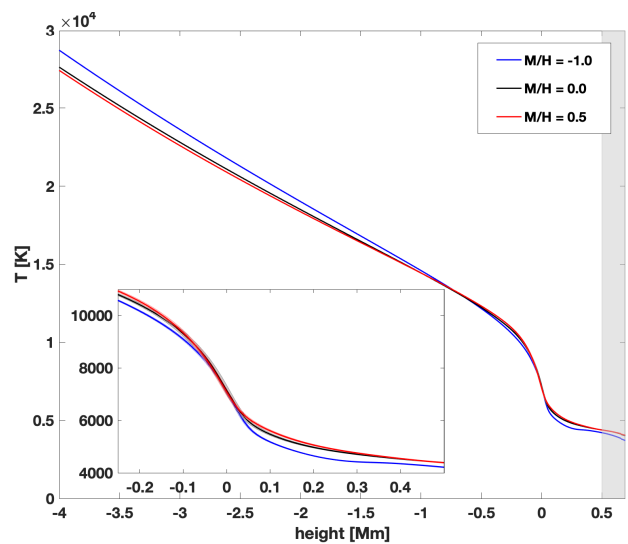}}
  \caption{Horizontally and time averaged temperature structure of the SSD runs with height for different metallicity values. The grey shaded area marks off the region that might be affected by the top boundary conditions. A more detailed discussion is in Sec.~\ref{subsec:sim_setup}.  }
  \label{fig:temp_with_height}
\end{figure}

\section{Results}
\label{Sec:results}

The one dimensional structures we show in this section are obtained by first averaging the three dimensional cubes in both horizontal directions. Subsequently, we take a time average over a series of snapshots over an interval of ten hours of solar simulation time (the typical number of snapshots is of order $10^3$). 


\subsection{Effect of metallicity on stratification of SSD and hydro runs}
\label{sebsec:stratification}

\begin{figure}
  \centering 
   {\includegraphics[width=1.0\linewidth]{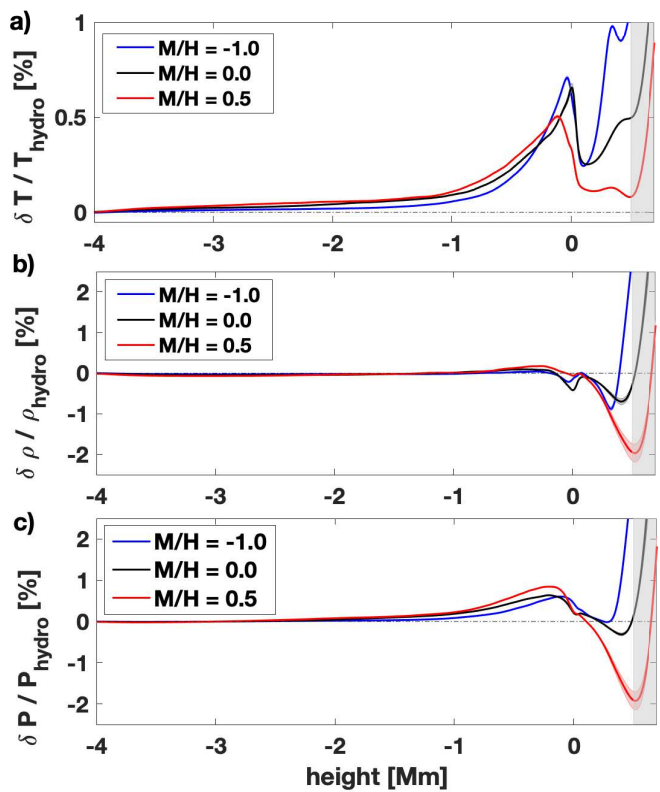}}
  \caption{Deviations in stratification from a pure hydro run to a SSD run. a) temperature deviations. b) density deviations. c) pressure deviations. The grey shaded area as in Fig.~\ref{fig:temp_with_height}.  }
  \label{fig:temp_ssd_to_h}
\end{figure}
In the following we consider the effect of metallicity on the stratification in the SSD simulations as well as the differences between SSD and pure hydrodynamic simulations. Overall, higher metallicity corresponds to greater opacity and leads to less efficient radiative transport of energy. 

We now focus on changes in temperature and density stratification. Fig.~\ref{fig:temp_with_height} shows the averaged 1D temperature with height, where we shifted the mean optical surface ($\tau_{500} = 1$) to $z = 0$ for each of the metallicity runs. In addition, one standard deviation, $\sigma$, over the temporal variation is plotted as a shaded area. If not otherwise stated, shifts of the mean optical surface and the one standard deviation are present in all figures of one dimensional structures. 
Since the mean effective temperature for the three metallicity runs is almost the same (see Table~\ref{table:01}), the temperature values in the vicinity of  $\tau_{500} = 1$ are similar. The low metallicity run (M/H = -1.0) shows a flatter temperature gradient just below the surface compared to the solar and the M/H = 0.5 run, but towards the deep layers in the convection zone the temperature shows a steeper increase compared to other metallicities. This can be explained by a somewhat lower pressure scale height for the ${\rm M/H =-1.0}$ run (see Appendix~\ref{app:stratification}).

For the low metallicity run the photospheric temperature averaged over the horizontal extent of the box and time decreases more strongly just above the surface  but then it flattens out while higher metallicity runs produce a slightly steeper temperature gradient. For a more detailed illustration of the changes in the temperature gradient see Fig.~\ref{afig:temp_with_p}. 
The differences in the temperature gradients around the optical surface and the significantly different temperature profiles in the photosphere with different metallicity are expected to noticeably change the emerging spectra. 
In particular, $\rm H^{-}$ is sensitive to small temperature changes and thus is expected to result in differences in the continuum opacities. Such differences will affect the energy distribution in the emergent intensities. In addition, temperature changes in the higher layers of the photosphere are expected to affect centre-to-limb variations. 

We note that while the dependencies of the atmospheric structures on M/H discussed above have  already been investigated in pure hydrodynamic simulations \citep[see, e.g.,][]{ludwig2009cfist, Magic_2013A&A} our calculations are the first which include SSD for stars with non-solar M/H. Along the same lines, previous studies \citep[see, e.g.][]{Magic_2013A&A, Magic2015A&A_paperIV} of spectra and limb darkening dependencies on metallicity have been performed without accounting for the effects of small-scale turbulent magnetic fields. We will perform a detailed analysis of these effects on emergent radiation for different metallicity values in a separate paper, while here we focus on changes of the atmospheric structure and dynamics introduced by the action of  SSD and the resulting small-scale turbulent magnetic field.

To pinpoint the effects of SSD,  we calculated the temperature deviations in the averaged structures obtained from the SSD run to the hydro run.   Fig.~\ref{fig:temp_ssd_to_h} shows that SSD heats the gas both below and above the optical surface. This implies a greater energy flux in the SSD calculations that leads to a slight increase of the effective temperature but mainly to excess heating in the upper photosphere, typical of magnetic flux concentrations (we remind that the bottom boundary conditions are the same for the HD and the SSD runs for each metallicity). 
This can be seen in Table~\ref{table:01} from the  effective temperature averaged over time together with one standard deviation. We note that  while the difference between the effective temperatures for the HD and the SSD run are within a standard deviation (i.e.~fluctuations in effective temperature from one time step to another due to convection are stronger than the permanent changes in $\rm T_{eff}$  due to a SSD), the values are significantly different (the error of the mean is much smaller than the standard deviation).  
At the surface, the increased heat flux in the SSD simulations can be explained by the magnetic regions being evacuated, which results in a density drop. The lower density in such magnetic regions allows for a more efficient radiative transport, because the opacity decreases. This in turn results in extra heating from hotter non-magnetic surroundings \citep[see][and references therein]{Solanki_et_al_2013}. Basically, the solar surface area is increased, so that more radiation can escape.

Below the optical surface, changes in pressure and density in the SSD run  (Fig.~\ref{fig:temp_ssd_to_h}) compared to the pure hydrodynamic simulation are rather insensitive to the metallicity and can be explained by the magnetic pressure and change of the turbulent pressure  \citep[see][for a more detailed discussion]{Bhatia_2022}.  
Higher up in the photosphere the changes in density and pressure depend significantly on the metallicity of the star. 


\subsection{Entropy and convection efficiency}
\label{subsec:entropy_convec}
\begin{figure}
  \centering 
    {\includegraphics[width=1.0\linewidth]{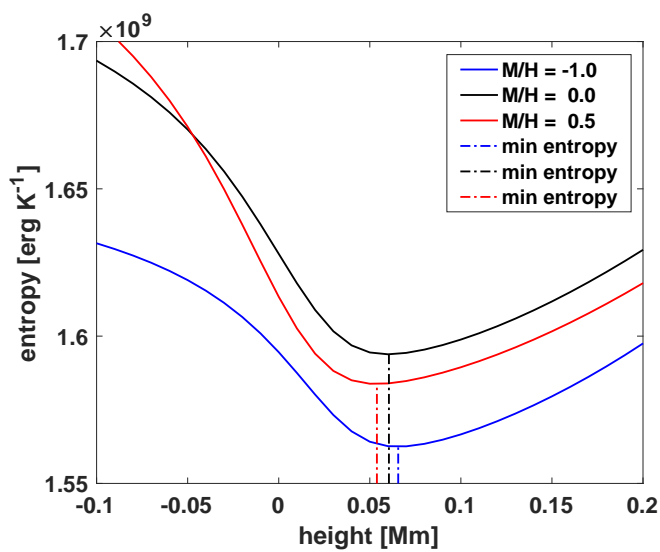}}
  \caption{Horizontally and time averaged entropy with height for the SSD simulations. The mean optical surface for all three runs is at $z=0$. The vertical dash-dotted lines indicate the position of the local entropy minima. }
  \label{fig:entropy}
\end{figure}

To quantify the effect of metallicity on convection  we first calculated the entropy profiles with height using the averaged density and internal energy profiles. Fig.~\ref{fig:entropy} shows the entropy minima and their positions for different metallicity values. The entropy minima are associated with the height at which the convection zone ends. It becomes evident that with decreasing metallicity the convective region extends further above the mean optical surface.

The entropy change, also called entropy jump, between the lower boundary (which corresponds to the constant entropy value present in the adiabatic convection zone) and at the entropy minima,  $\Delta S$,  can be used as an indicator of the convection efficiency  \citep{Magic_2013A&A}. 
Table.~\ref{table:01} shows that the effect of metallicity on the $\Delta S$ is significant, resulting in a greater entropy jump for greater metallicities.  At the same time the difference between a pure hydrodynamic setup and one including SSD action is very small. 
A strong dependence of the entropy jump on metallicity seen in our simulations agrees with previous pure hydrodynamic studies, where it was shown that with decreasing metallicity the convection efficiency increases \citep{Magic_2013A&A}. A more efficient convection is expected to lead to stronger magnetic field generation, which we discuss further below.

\begin{table}
\setlength\tabcolsep{5pt}
\renewcommand{\arraystretch}{1.5}
\caption{Mean effective temperature and analysis of the entropy profiles. Time-averaged mean effective temperature calculated from the bolometric intensity that is written out at each iteration step over 10 hours together with one standard deviation in the temporal fluctuations.  Entropy change between the local minima and the bottom boundary for different metallicities in the SSD simulation, $\Delta$ S. Ratio of the entropy change in the SSD  to the hydro run. Height position of local entropy minima compared to optical surface.}              
\label{table:01}      
\centering                                     
\begin{tabular}{|c | c| c| c|  }         
\hline\hline                       
M/H [dex]       & -1.0  & 0.0 & 0.5   \\   
\hline\hline                               
 SSD $\rm T_{eff} [K]$ & 5772$\pm 5$   & 5787 $\pm 9$  & 5782 $\pm 11$ \\
 \hline\hline 
HD $\rm T_{eff} [K]$ &  5767 $\pm 4$  &5782 $\pm 8$ & 5776 $\pm 11$ \\
 \hline\hline 
 $\Delta \rm S_{SSD}$  [erg $\rm K^{-1}$] & 1.1732 $10^8$  & 1.8133 $10^8$   & 2.3756 $10^8$\\
 \hline\hline                                 
 $\Delta \rm S_{SSD} $/  $\Delta \rm S_{hydro} $  & 0.9970  & 0.9971 & 0.9981   \\
  \hline\hline                                 
 $\delta$ h [Mm] & 0.0656  &  0.0605 & 0.0539  \\
\hline\hline    
\end{tabular}
\end{table}

\subsection{Velocities, vertical momentum,   \& turbulence }
\label{subsec:velocities}

\begin{figure}
   \centering 
    {\includegraphics[width=1.0\linewidth]{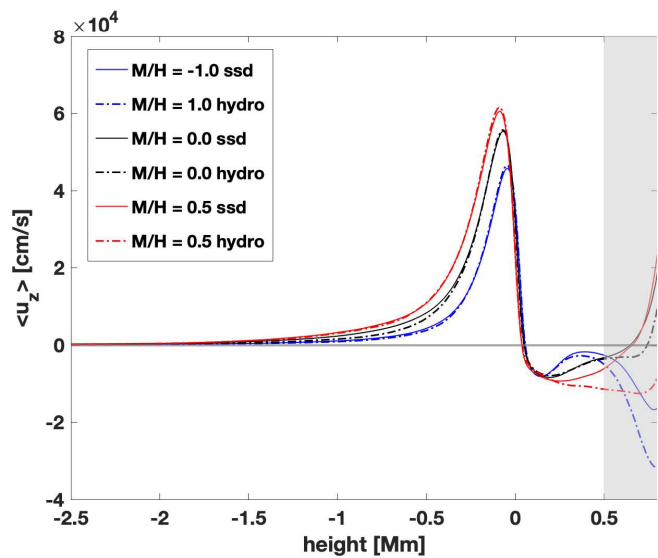}}
  \caption{Horizontally and time averaged vertical velocity for different metallicity values. The grey shaded area as in Fig.~\ref{fig:temp_with_height}.  }
  \label{fig:averaged_vertical_vel}
\end{figure}

To investigate how metallicity and SSD affects the dynamics we plot various quantities. We start by showing the horizontally and time averaged vertical velocity, $\rm <u_z>$, in Fig.~\ref{fig:averaged_vertical_vel}.
Below the optical surface, $\rm < u_z >$ increases for all metallicities and peaks close to the surface, but at different heights and reaching different values (highest peak for largest metallicity value) with metallicity.
For all metallicities a steep drop in $< u_z >$ occurs at the surface followed by very distinct behaviour above the surface. Moreover, above the surface  the  $\rm < u_z >$ is significantly affected by the SSD. 

\begin{figure}
   \centering 
{\includegraphics[width=1.0\linewidth]{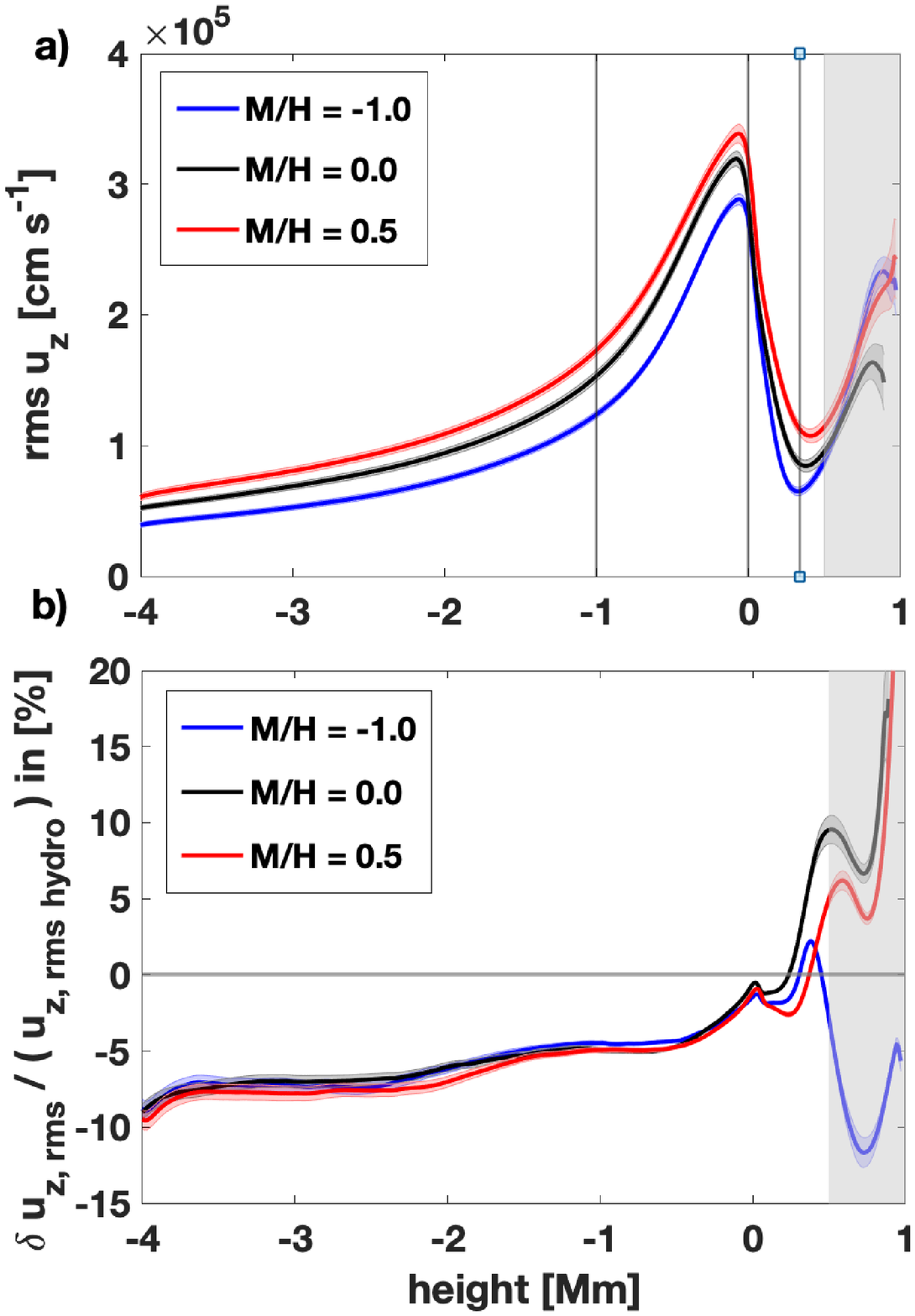}}
    
  \caption{a) Root-mean-square vertical velocities,  $\rm u_{z,rms}$,  for the SSD run and b) the deviation between the SSD  and the hydro run for different metallicity values. The grey shaded area as in Fig.~\ref{fig:temp_with_height}. The vertical lines denote heights at which the velocity distribution is obtained.}
  \label{fig:vertical_ssd}
\end{figure}

 In Fig.~\ref{fig:vertical_ssd} we show the averaged rms vertical velocity, $\rm u_{z,rms}$, and its deviation in the SSD compared to a pure hydro simulation. The root-mean-square vertical velocities can inform us more about the turbulent motion.
 The rms vertical velocity gradually increases with height in the convection zone and reaches a peak just below the optical surface (see Fig.~\ref{fig:vertical_ssd}a). 
Greater metallicity values lead to higher rms vertical  velocities at almost all heights.  In addition, the SSD significantly affects $\rm u_{z,rms}$  both above and below the surface (see Fig.~\ref{fig:vertical_ssd}b). Below the surface a decreased rms vertical velocity (up to 10\%) in the SSD run compared to the hydro run is observed.
While the deviations are insensitive to metallicity below the optical surface, above the surface they  significantly differ for the three metallicities. For M/H = -1.0 the vertical rms velocity is suppressed also in the photosphere, while for the higher metallicity values it is enhanced.   We note that difference in vertical velocity profiles will show itself in profiles of spectral lines. We will perform the line synthesis and discuss this effect in detail in the forthcoming publication. 

\begin{figure}
   \centering 
    {\includegraphics[width=1.0\linewidth]{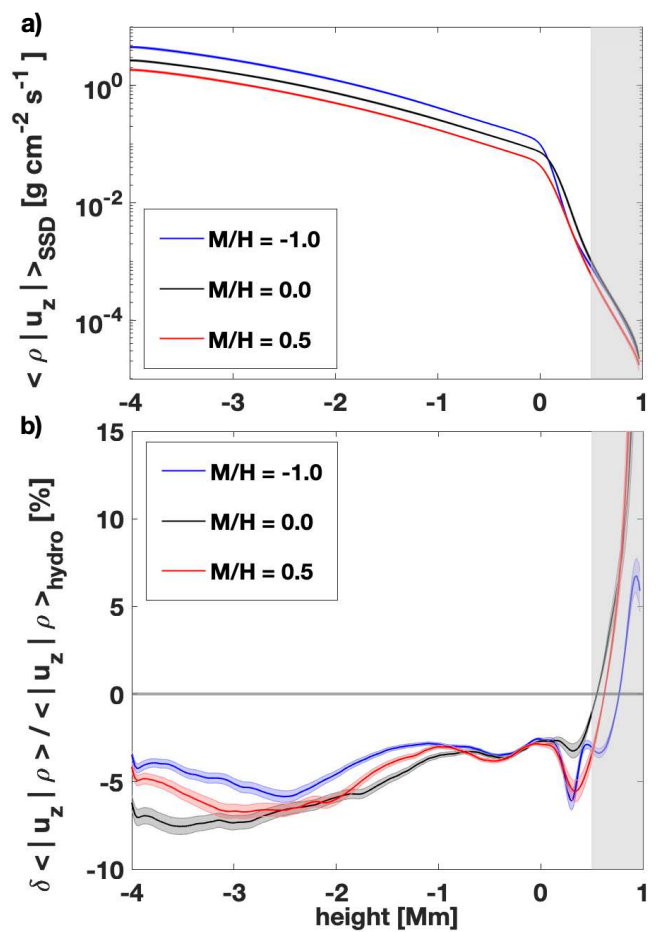}}
  \caption{Vertical momentum (top panel) for different M/H and ratio of vertical momentum in the SSD to the hydro simulation (bottom panel). The grey shaded area as in Fig.~\ref{fig:temp_with_height}.  }
  \label{fig:vmom_ssd}
\end{figure}

Another important characteristic of convection is the vertical momentum transport. First, it is useful to calculate the mean absolute vertical velocity, $|u_z|$, weighted with density, $<\rho |u_z|>$, as shown in Fig.~\ref{fig:vmom_ssd} a). This quantity is proportional to the vertical momentum transport.
With decreasing metallicity, $<\rho |u_z|>$ increases significantly, which indicates that also the vertical momentum transport for lower metallicity stars is stronger at the optical surface and below it.  

Comparing the SSD run to the pure hydrodynamic run, the momentum transport is suppressed by the small-scale turbulent magnetic fields.
Indeed, Fig.~\ref{fig:vmom_ssd} b shows that throughout the convection zone the vertical momentum transport is decreased by about 5\%  on average for the solar metallicity, by contrast, in the low metallicity run the decrease is slightly smaller.   
A similar decrease up to 10\% was also observed for the rms vertical velocity in Fig.~\ref{fig:vertical_ssd} b). This indicates that the observed decrease in the vertical momentum transport below the surface is mainly due to suppression of convection and thus vertical motion by magnetic fields. 
%

\begin{figure*}
   \centering 
      {\includegraphics[width=1.0\linewidth]{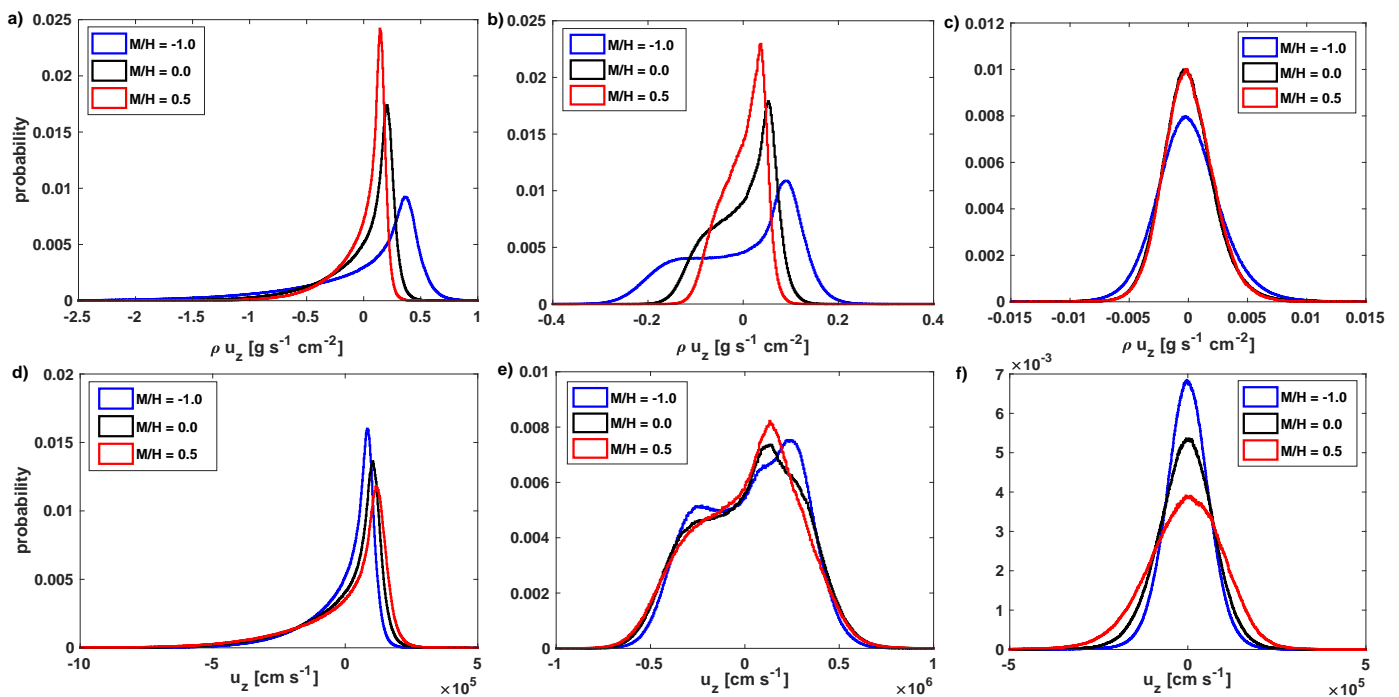}}
 \caption{Histograms of vertical momentum density top panels  and vertical velocity bottom panels obtained in SSD simulations. Left column is at 1Mm below the optical surface, middle column shows the distributions at the optical surface and right column at about 300 km above the optical surface. 
  }
  \label{fig:hist_vertical_mom}
\end{figure*}

Second, to investigate the dynamics of the mixing that occurs, we calculated histograms of the vertical momentum density and vertical velocity for three layers around and at the optical surface as indicated by the vertical lines in Fig.~\ref{fig:vertical_ssd} a. The top layer was chosen, because the rms vertical velocity shows a local minima there, while the layer below the surface is just below the inflection point. 
The distribution of the vertical momentum density (shown in Fig.~\ref{fig:hist_vertical_mom} a-c) above the surface is symmetric and peaks at zero. Whereas at the surface and below it shows clearly the asymmetry of the up- and down-flows. For all three layers the distribution widens with decreasing metallicity (most prominent at the optical surface). At the surface a decrease in metallicity results in a more pronounced shoulder corresponding to the down-flows.

When comparing the vertical velocity distributions at these three layers (Fig.~\ref{fig:hist_vertical_mom} d - f), the distributions are asymmetric below the surface and symmetric above. On the contrary, there is no significant broadening with different metallicity, even more above the surface the distributions become wider with increasing metallicity.   
 The change of the down-flow shoulder is also clearly visible in the vertical velocity distributions (see Fig.~\ref{fig:hist_vertical_mom}e). Consequently, one can expect that such changes in velocity distribution manifest themselves in the profiles of spectral lines.  

\begin{figure}
   \centering 
    {\includegraphics[width=1.0\linewidth]{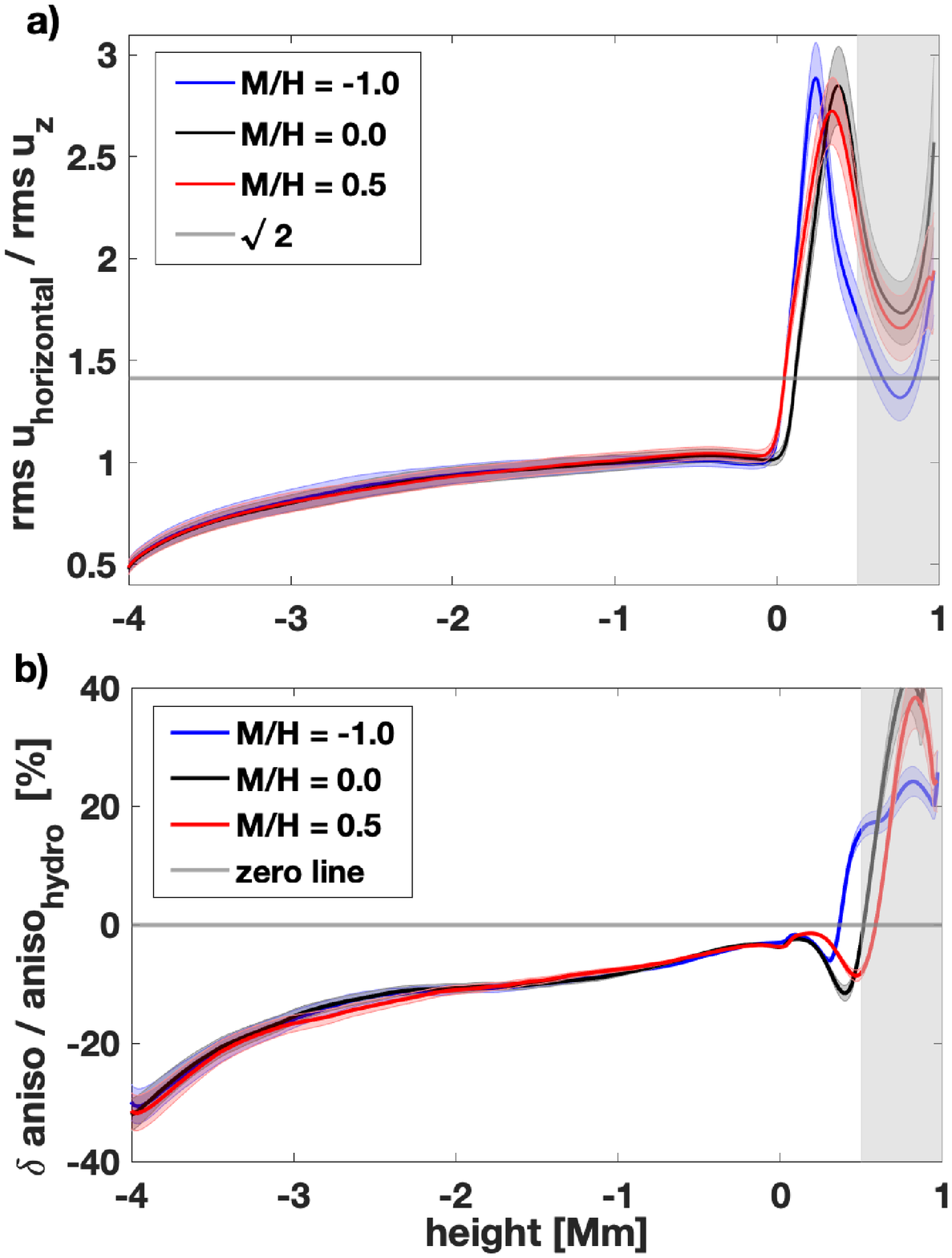}}
  \caption{Ratio of horizontally averaged rms horizontal to rms vertical  velocity with different metallicity in (a). The grey dotted line indicates the ratio for an isotropic case. b) The deviation of this ratio in the SSD run compared to the hydro run. The grey shaded area as in Fig.~\ref{fig:temp_with_height}. }
  \label{fig:anisotropi_ssd}
\end{figure}

The anisotropy of the velocity is plotted in Fig.~\ref{fig:anisotropi_ssd} a) for the SSD simulations. It shows that velocities in the convective zone are predominantly vertical, i.e.~the ratio of rms horizontal to rms vertical velocity is below $\sqrt{2}$. Interestingly, below and at the optical surface this ratio does not significantly deviate for different metallicity values. This indicates that the morphology of the convection cells and granulation is only weakly affected by metallicity.  
On the contrary, in the photosphere the velocity becomes predominantly horizontal, with vertical velocity decreasing just above the surface (see Fig.~\ref{fig:averaged_vertical_vel}).
In the mid photosphere the anisotropy of rms velocities shows a rather complex behaviour with metallicity. When comparing the anisotropy of the velocities in the SSD simulation to the hydro simulation (see Fig.~\ref{fig:anisotropi_ssd} b) the anisotropy is significantly stronger for the SSD run below and above the optical surface. In addition, the change from hydro to SSD is strongly affected by metallicity above the optical surface, while below the surface only a weak dependence on metallicity can be observed. Overall, the difference in velocities introduced by metallicity and SSD  is expected to affect the profiles of spectral lines.

To get a more detailed picture of the characteristics of the turbulent motion we looked at the kinetic energy power spectra at different heights. For that we consider a  layer at the optical surface ($\tau_{500} =1$) and further layers in 1Mm steps below the surface layer.  The calculated power spectra are averaged in these four layers over 10 hours of stellar time (ca.~30 cubes were used for the averages).

\begin{figure*}
   \centering 
    {\includegraphics[width=1.0\linewidth]{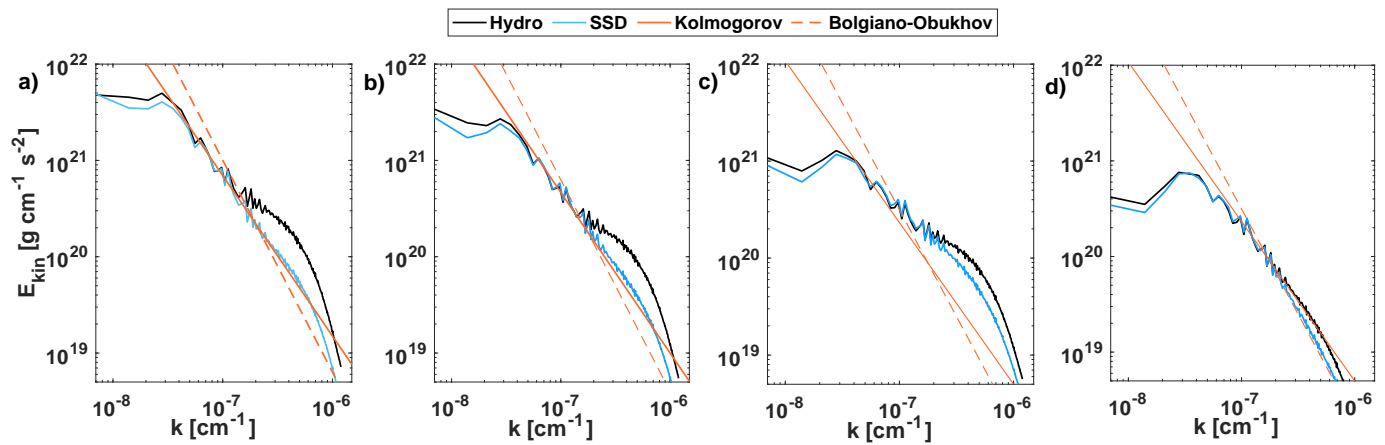}}
  \caption{Kinetic energy power-spectra for four layers at different heights for the SSD and hydro runs with metallicity M/H = 0.0. Additionally, the Kolmogorov and the Bolgiano-Obukhov scaling is plotted. a) at -2.5 Mm below the optical surface b) at -1.5 Mm below the optical surface, c) at -0.5 Mm below the optical surface, d) at the optical surface.  }
  \label{fig:e_spectra_ssd_vs_hydro}
\end{figure*}

Fig.~\ref{fig:e_spectra_ssd_vs_hydro} shows the kinetic energy power spectra for the SSD run and the pure hydrodynamic run. One can see that the overall energy decreases with height (compare various panels of Fig.~\ref{fig:e_spectra_ssd_vs_hydro}), which is expected as the density changes several orders of magnitude between the bottom and surface of the simulated region.
Moreover, the distribution of energy over length scales changes with height (which is clearly visible by comparing  the gradient of the power spectra). For the SSD simulation, in the two deeper layers the kinetic energy power spectra shows rather an agreement with the theoretical Kolmogorov scaling. However, in the upper layers, in particular, at the optical surface, for wavenumbers k greater than $1.0\times 10^{-7} \rm cm^{-1}$ an agreement with the Bolgiano-Obukhov scale is present. %
Interestingly, in the pure hydrodynamic run, there is clearly more kinetic energy for wavenumbers k greater than $1.5\times 10^{-7} \rm cm^{-1}$, which corresponds to length scales of about 450km,  compared to the SSD run.  Two effects play a role: on the one hand kinetic energy is converted to magnetic energy and on the other hand magnetic fields back-react and suppress motion. Since the difference increases with depth, it indicates that both effects become stronger in the lower layers. 
%
Metallicity does not lead to a significant effect on the shape of the kinetic energy power spectra at any of the heights (see Fig.~\ref{afig:e_spectra_with_layers} and more detailed discussion in Appendix.~\ref{app:energy_spec}). Only the overall energy is affected by metallicity, whereby in low metallicity runs the kinetic energy is greater. Thus, we expect greater magnetic field strengths in the low metallicity simulation, which will be discussed in the next Section.

\subsection{Magnetic fields from SSD}
\label{subsec:magneticfields}
Small-scale turbulent magnetic fields are observed in the vicinity of the optical surface on the Sun. Such magnetic fields could be explained by a small-scale turbulent dynamo action in the upper convection zone \citep{Schuessler_voegler2008, rempel_2014}. While the small-scale turbulent magnetic fields generated by an SSD were studied for the solar case,  here we aim to investigate how these magnetic fields are affected by the metallicity of a star. 

\begin{figure}
   \centering 
   {\includegraphics[width=1.0\linewidth]{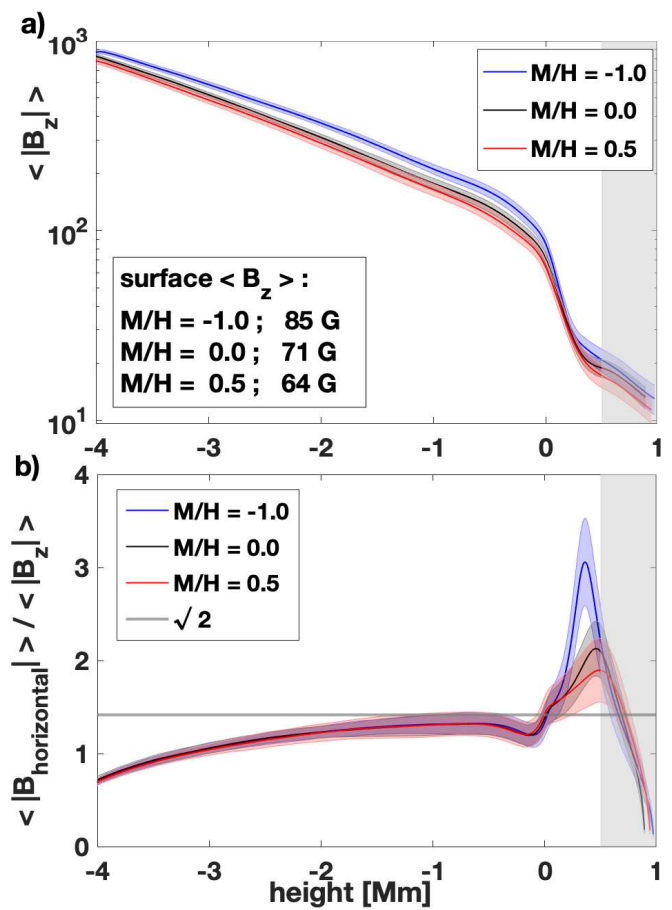}}
    
  \caption{Vertical magnetic field (a), and ratio of horizontal to vertical magnetic field (b) for the SSD simulation. The grey shaded area as in Fig.~\ref{fig:temp_with_height}. }
  \label{fig:vertical_b_ratio_ssd}
\end{figure}

We first focus on analysing the mean absolute vertical magnetic field strength as a function of geometrical height, shown in Fig.~\ref{fig:vertical_b_ratio_ssd}. It becomes evident that for low metallicity stars, the vertical magnetic field strength is on average significantly higher in the upper convection zone compared to the solar and the $\rm M/H=0.5$ run. This is expected since the magnetic field generation occurs due to the turbulent motion in the convective layers. Thus, stronger magnetic fields for the $\rm M/H = -1$ metallicity are expected due to the increase of the kinetic energy with decreasing metallicity (see Fig.~\ref{afig:spectra_with_layers} and a discussion on the efficiency in Appendix~\ref{app:energies}). 

Magnetic fields generated in the convective zone emerge at the surface and are transported upwards. Therefore, closer to the optical surface there is a steep drop in the averaged magnetic field strength leading to similar mean absolute vertical magnetic field strength values at the optical surface for all three metallicities that we consider. For the solar metallicity, the mean value of the absolute vertical magnetic field at the optical surface of 71G agrees very well with previous theoretical work \citep{rempel_2014,Khomenko_2017A&Aetal,Rempel_2018ApJ}  as well as observations \citep{Trujillo_Bueno_2004Natur, Danilovic_2010A&A, sanchez_2011ASPC, Shapiro_2011A&A}. A slightly higher mean absolute magnetic field of 85G is present for M/H = -1.0 and a slightly lower one of 64G for M/H = 0.5. Interestingly, a strong gradient of the field in the photosphere might be an important factor contributing to the large spread of magnetic field values determined using the Hanle effect analysis of different spectral lines \citep[see, e.g.,][and references therein]{Lucia2011, Ivan2012}. 

Focusing on the anisotropy of magnetic fields, shown in Fig.~\ref{fig:vertical_b_ratio_ssd}b, we find that below the surface the magnetic fields are almost isotropic (independently of the metallicity). On the contrary, above the optical surface predominant horizontal magnetic fields are present and the anisotropy of the magnetic fields shows a dependency on the metallicity. Our findings for  solar metallicity are consistent with \citet{Schuessler_voegler2008}, who also showed that an SSD leads to predominantly horizontal fields in solar middle photosphere. 

The layers above and at the optical surface are of particular interest, as most observational techniques measure the magnetic field strength at different heights above the optical surface. Thus, we show the distribution of the vertical magnetic field strength in three layers, where two of them are above the optical surface and the third one at the optical surface.

\begin{figure*}
   \centering 
    {\includegraphics[width=1.0\linewidth]{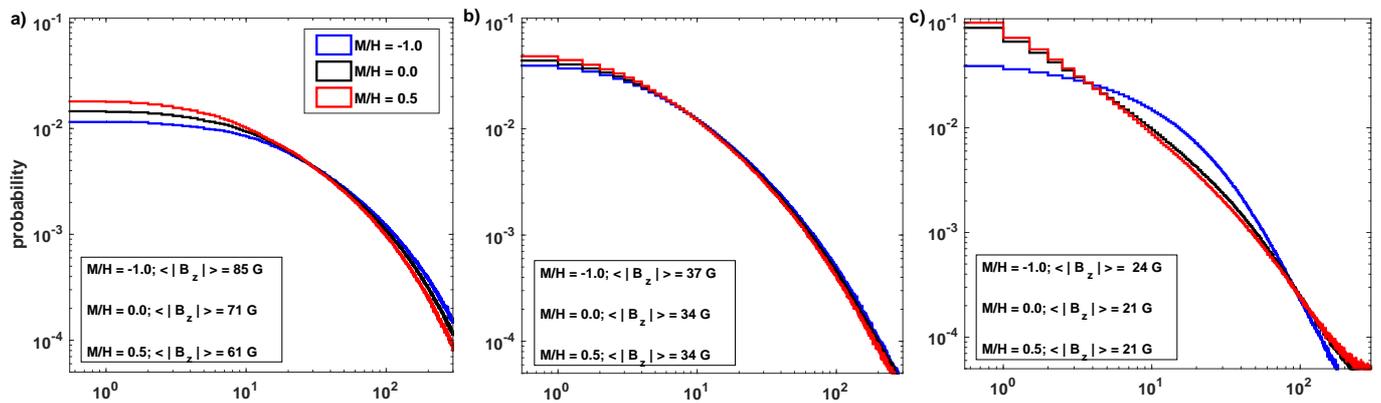}}\\
  \caption{Histogram of vertical magnetic field strength, $|B_z| $,  distribution for layer three layers. a) at the optical surface, b) at 170km above the optical surface, and c) 340km above the optical surface. }
  \label{fig:vertical_mag_dist}
\end{figure*}

At the optical surface not only the mean value of the absolute vertical magnetic field decreases with metallicity (see Fig.~\ref{fig:vertical_b_ratio_ssd}), but also the area covered by the concentrations of relatively strong field (see the tails of distributions plotted in Fig.~\ref{fig:vertical_mag_dist} a). Furthermore, at the surface the maximal values reached are significantly different for the three metallicities ranging from roughly 1800G for M/H = 0.5 to roughly 2500G for M/H = -1.0. 

At about 170km above the surface, shown in Fig.~\ref{fig:vertical_mag_dist} b), the distribution of the vertical magnetic field shows a weaker metallicity dependence compared to the optical surface (the mean absolute vertical field is roughly the same for all three metallicities). In general a flatter and wider distribution occurs with decreasing metallicity. Except for M/H = -1.0  further up at 340km above the optical surface (see Fig.~\ref{fig:vertical_mag_dist} c), where  a significantly different distribution is present.

\subsection{Effect of metallicity on surface structure}
\label{subsec:2D_illustrations}
\begin{figure*}
  \centering 
   {\includegraphics[width=1.0\linewidth]{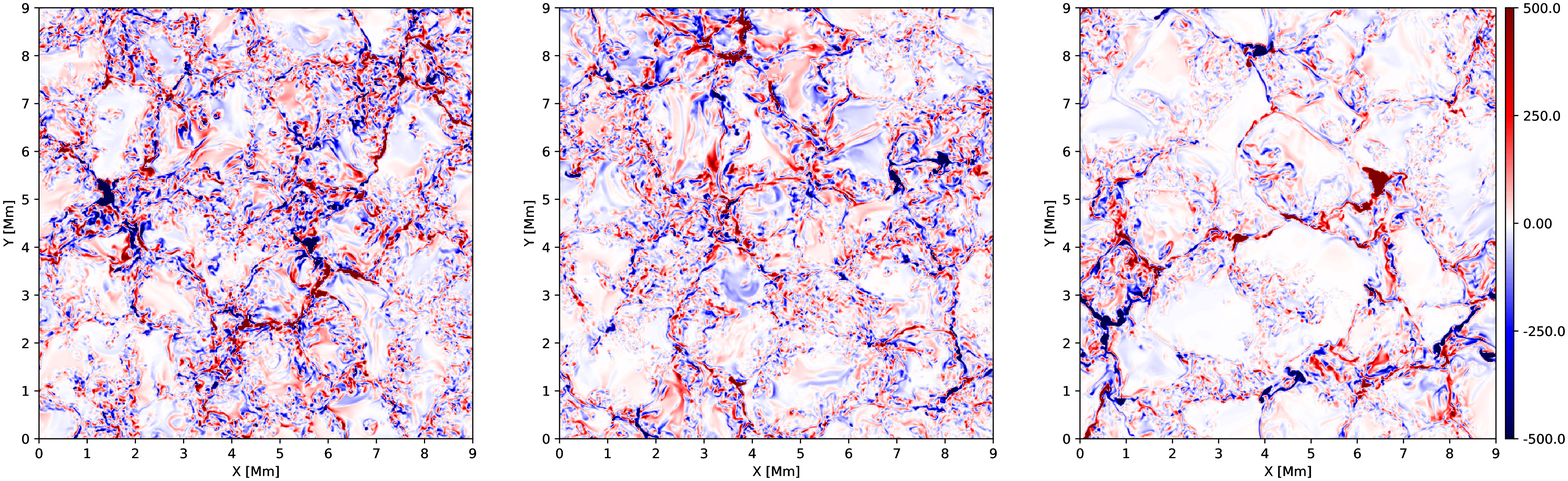}}\\
   {\includegraphics[width=1.0\linewidth]{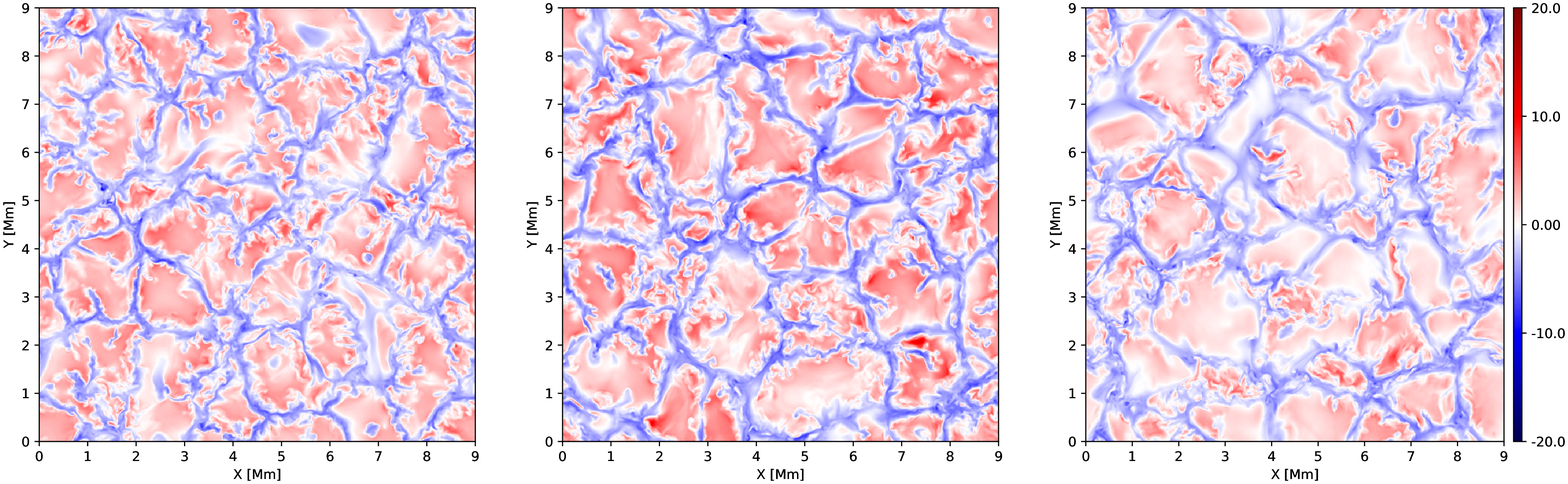}}\\
   {\includegraphics[width=1.0\linewidth]{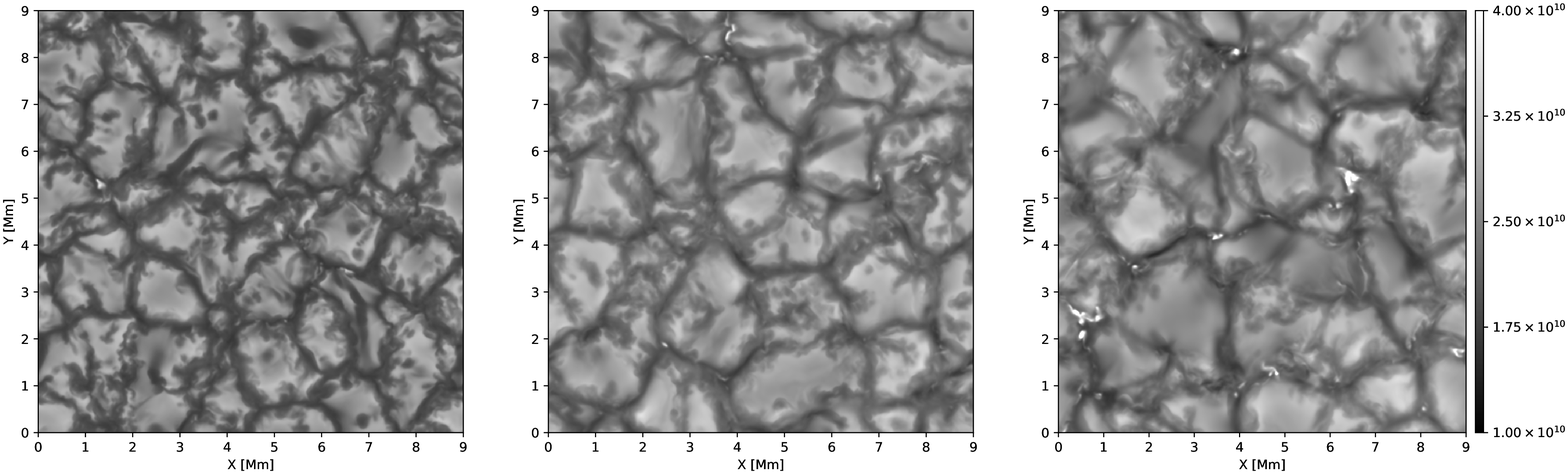}}\\
  \caption{ Vertical magnetic field, $B_z$, vertical velocity, $\rm u_z$,  at $\tau_{500} = 1$  and bolometric intensity at disc centre for different metallicity values (from left to right M/H = -1.0, M/H = 0.0, M/H = 0.5). Top panels: $B_z$ in units of [G],   Middle panels: $\rm u_z$ in units of [$\rm km$ $\rm s^{-1}$] Bottom panel: bolometric intensity in units of [$\rm erg$ $\rm cm^{-2}$ $\rm s^{-1}$].}
  \label{fig:visit_plots}
\end{figure*}
To illustrate the effect of metallicity on the surface structure,  Fig.~\ref{fig:visit_plots} shows example 2D plots of the vertical magnetic field, the vertical velocity at the optical surface, and the bolometric intensity at disc centre. The structure of the vertical magnetic fields look very similar for the three metallicities. This is not surprising, as Fig.~\ref{fig:vertical_b_ratio_ssd} showed that the strength of the vertical fields on average is very close to each other for the three metallicities around the optical surface. For the vertical velocities stronger up- and down-flows occur for the solar metallicity, but no clear trend can be identified from the snapshots. 
Finally, the bolometric intensity clearly shows that the intensity contrasts between inter granule lanes and granules are different. More small-scale structures become visible for  M/H = -1.0. Moreover,  the most prominent bright points are present in the M/H = 0.5 run. This shows that the effect of metallicity on the radiative transfer is complex and requires detailed calculations.


\section{Summary \& Discussion}
\label{Sec:discussion}
The solar surface, even in its quietest state,  is covered by small-scale turbulent magnetic fields of mixed polarity \citep[][]{Khomenko2003A&A,sveta2004, Trujillo2004Natur,Lites2004ApJ,Stenflo2013}, which could be explained by a SSD  driven by near-surface convection \citep{Voegler_Schuessler_2007A&A, Schuessler_voegler2008, rempel_2014}. Until now the  SSD simulations focused on the Sun and achieved good agreement with observations \citep{Danilovic_2010A&A, Danilovic2016A&A}. However, we have been lacking SSD simulations for other cool stars. Here, we simulated for the first time the action of SSD in stars with non-solar metallicity values  and investigated  the effect of metallicity on the stratification, convection and magnetic field generation and distribution in SSD simulations. 

Our findings for the stratification and convection characteristics for purely  hydrodynamic calculations are in agreement with previous studies \citep[e.g.,  by][]{Magic_2013A&A}. The small-scale turbulent dynamo calculations show similar trends with metallicity, but at the same time differ noticeably from the pure hydrodynamic calculations. In particular, momentum transport  significantly changes for the SSD run in comparison with pure hyrodynamic simulations and furthermore depend on metallicity throughout the convection zone and in the lower photosphere. Also vertical velocities differ noticeably in the lower photosphere with metallicity.

Moreover, we showed that the magnetic fields generated by a SSD operating in stars of different metallicity have different properties. Generally,  low metallicity stars show vertical magnetic fields of greater strength compared to stars with higher metallicity. In addition, the ratio of vertical  to  horizontal magnetic field strength in the photosphere changes with metallicity.

The found differences in stratification and velocities are expected to affect observable quantities, such as overall spectral emission,  contrasts of granules, the centre-to-limb variations, and line profiles. 

So far we analysed intrinsic properties of the 3D simulations, where this study focused on different metallicities for solar effective temperature and a separate study investigated different spectral types (F3V, G2V, K0V, and M0V) \citep{Bhatia_2022}. 
However, most of the quantities calculated in 3D simulations can not be observed directly even on the Sun.  To understand how metallicity and small-scale turbulent magnetic fields affect observable quantities, a comprehensive radiative transfer calculations are needed. Such calculations, using the MPS-ATLAS code \citep{mps-atlas-w2021},  are currently on the way and will be presented in the forthcoming publication.

\begin{acknowledgements}
This work has received funding from the European Research Council (ERC) under the European Union's Horizon 2020 research and innovation programme (grant agreement No. 715947). 
\end{acknowledgements}

\bibliographystyle{aa} %
\bibliography{refer} %

\newpage

\begin{appendix}
\section{Stratification}
\label{app:stratification}
\begin{figure}
  \centering 
    {\includegraphics[width=1.0\linewidth]{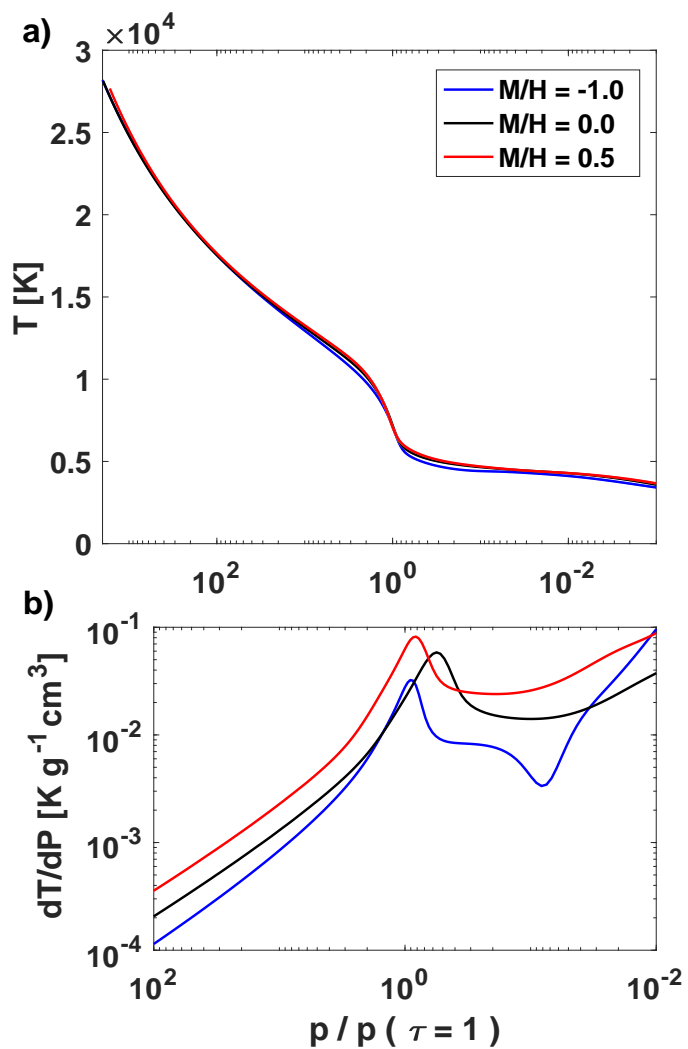}}
  \caption{Horizontally and time averaged temperature structure and gradient for the SSD simulations versus pressure for different metallicity values. a) temperature  b) gradient of temperature with respect to pressure. }
  \label{afig:temp_with_p}
\end{figure}

Metallicity directly affects the density, and thus the pressure of the simulated plasma. Thus, another useful scale to illustrate the change in  stratification is the pressure. For that we calculate the horizontally and time averaged pressure, $p(z)$, and normalise it by the pressure value at the optical surface $p(z = 0)$. The averages are taken at the same geometrical heights (not on iso-pressure surfaces). Fig.~\ref{afig:temp_with_p} shows the temperature structures and the temperature gradients with respect to pressure for different metallicity values. 

The temperature gradient, shown in Fig.~\ref{afig:temp_with_p} b, is smaller (almost at all heights) for lower metallicities so that lower metallicity allows for more efficient radiative transfer and thus for a more efficient energy transport. A local maximum in the temperature gradient close to the optical surface occurs for all three metallicities. This maximum indicates the threshold where radiative transfer becomes more efficient compared to the convective transport of energy. We note that the significantly different temperature gradients  affect the radiative transfer differently for different frequency ranges of the spectra, leading to different spectral distribution of emitted radiation. 
\section{Energy spectra and typical length scale}
\label{app:energy_spec}
\begin{figure}
  \centering 
    {\includegraphics[width=1.0\linewidth]{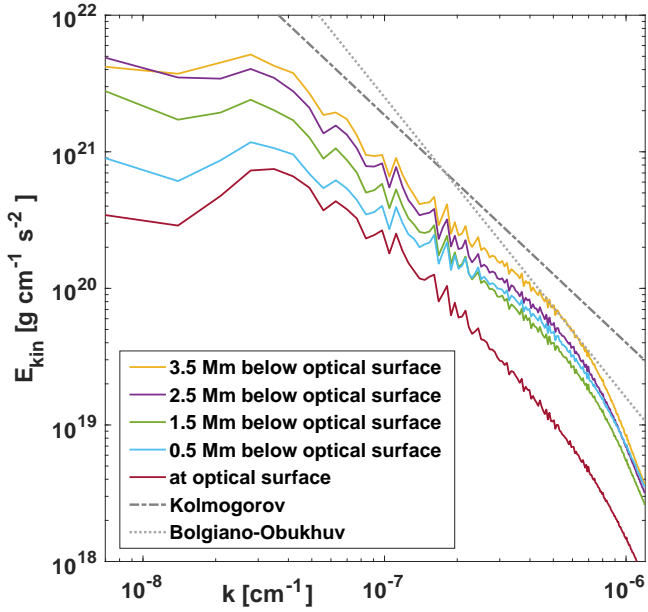}}
  \caption{Kinetic energy power-spectra at different heights for the SSD case with M/H = 0.0. Additionally, the Kolmogorov and the Bolgiano-Obukhov scaling is plotted.}
  \label{afig:spectra_with_layers}
\end{figure}

The kinetic energy power-spectra are obtained by calculating 
\begin{eqnarray}
\label{eq:power-spec}
    E(k) &  =  & \frac{1}{4} \sum_{}^{}  \mathbf{\hat{u}}(k_x,k_y, z)  \hat{\rho \mathbf{u}}^{*} (k_x, k_y,z ) \, \nonumber   \\
    &   & + \mathbf{\hat{u}}^{*}(k_x, k_y, z)  \hat{\rho \mathbf{u}} (k_x, k_y,z),
\end{eqnarray}
where $k^2 = k_x^2 + k_y^2$, the $\hat{ }$ indicates a Fourier-transform, the $ ^{*}$ a complex conjugation, $\mathbf{u}$ is the 3D velocity field, and z denotes the vertical height of the layer.

Fig.~\ref{afig:spectra_with_layers} shows the kinetic energy power-spectra for different layers in the SSD run with solar metallicity. 
The calculated power spectra are averaged in these five layers  with time over at least 30 cubes, which corresponds to at least 10 hours of stellar time. The overall kinetic energy in the layers decreases with height, which is expected as the density changes several orders of magnitude between the bottom and surface of the simulated region.  Moreover, the slope of the power spectra changes. It becomes steeper for wavenumber smaller than $\rm 5 \times 10^{-7} cm^{-2}$ (which corresponds to a length scale larger than 1250 km). 

\begin{figure*}
   \centering 
    {\includegraphics[width=1.0\linewidth]{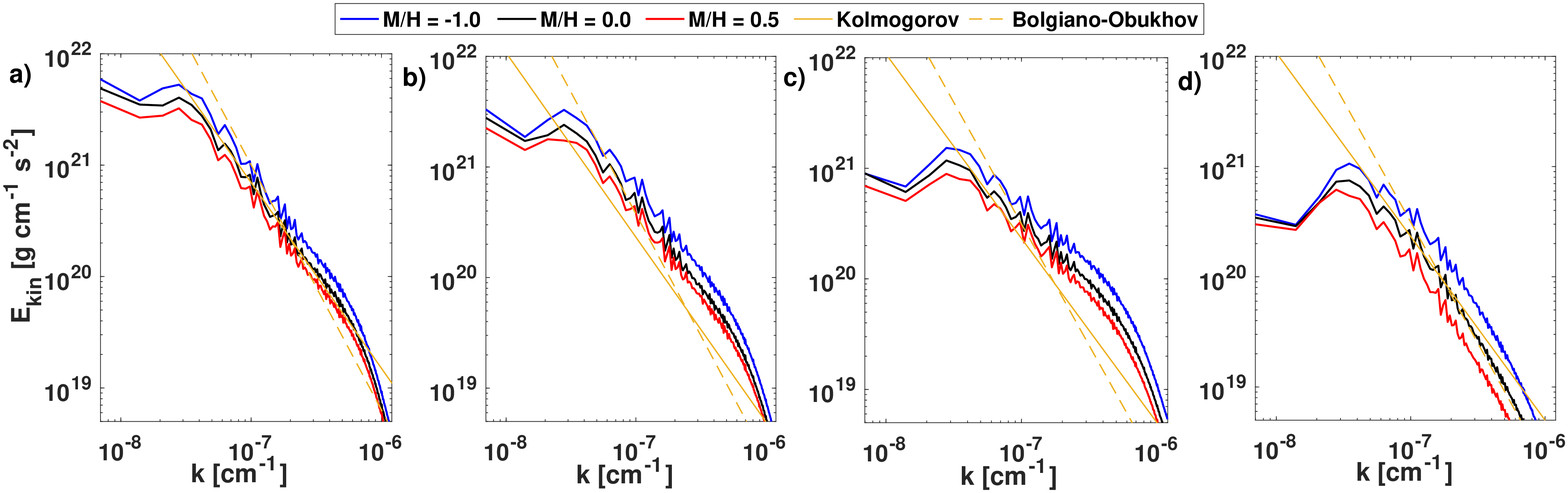}}
  \caption{Kinetic energy power-spectra at different heights for different metallicity values of the SSD simulation. Additionally, the Kolmogorov and the Bolgiano-Obukhov scaling is plotted. a) layer at -2.5Mm below the optical surface b) at -1.5 Mm below the optical surface, c) at -0.5 Mm below the optical surface, d) layer at the optical surface. }
  \label{afig:e_spectra_with_layers}
\end{figure*}

In Fig.~\ref{afig:e_spectra_with_layers} for each of the first four layers the kinetic energy spectra are shown with different metallicity. It becomes evident that while the overall kinetic energy is smallest for the $\rm M/H = 0.5 $ the distribution of the kinetic energy over the length-scales is not significantly affected. 


Finally, to estimate a typical turbulence length-scale,  we assume that the turbulence in the horizontal direction is isotropic and employ the integral-length-scale formulation
\begin{equation}
\label{eq:integral_l}
    l_{typ} =  \frac{\sum_k  2\pi  E(k)/ k}{\sum_k E(k) }. 
\end{equation}
This quantity is associated with the size of the turbulent eddies that have most of the energy \citep{batchelor1953theory}. It can also interpreted as the sizes of the most predominant the convective cells. 
For the analysis we calculate $l_{typ}$ at each layer in the simulated domain but take the temporal average over the whole 10 hours.

\begin{figure}
   \centering 
    {\includegraphics[width=1.0\linewidth]{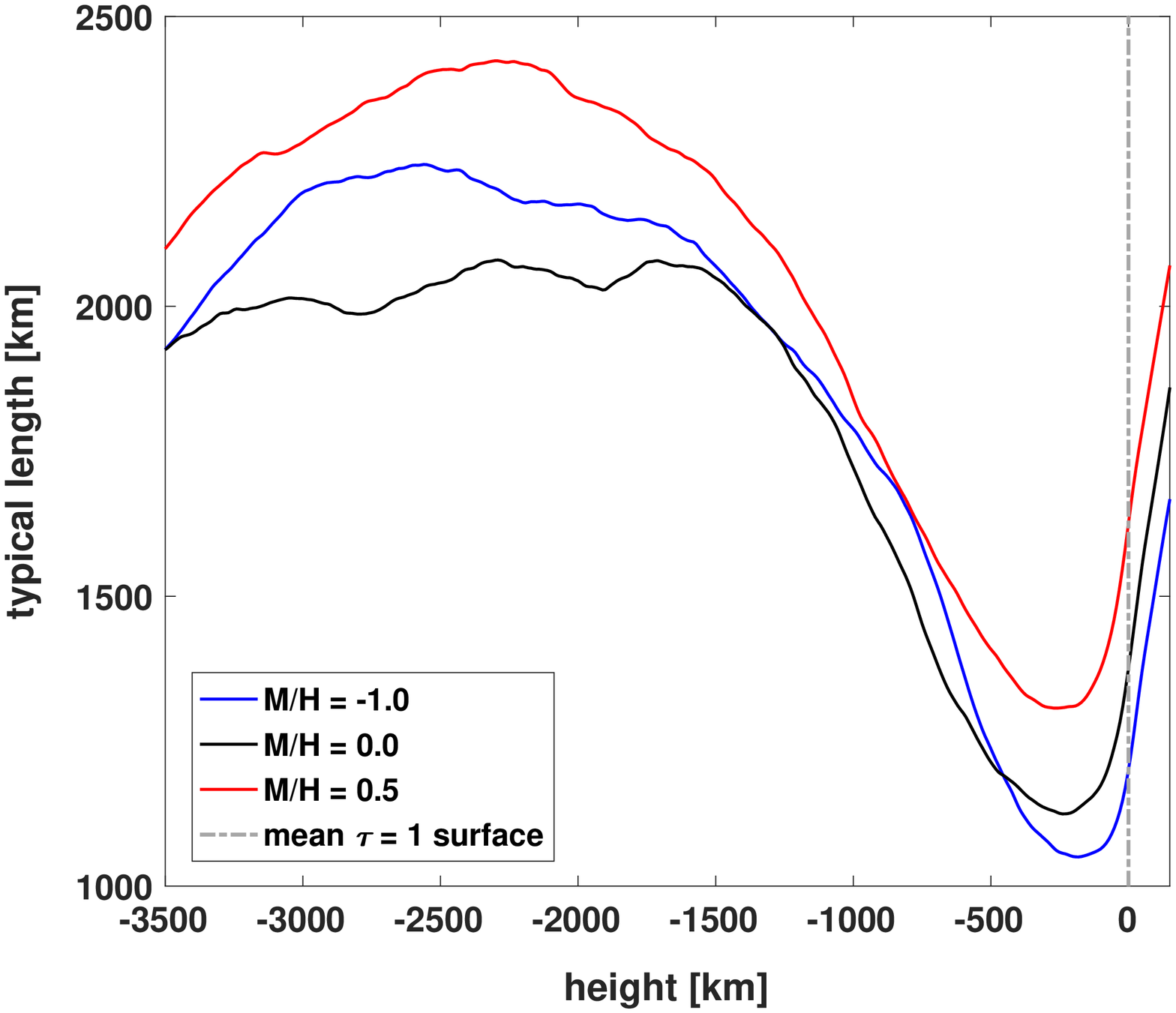}}\\
  \caption{Typical integral-length-scale as defined in equation \eqref{eq:integral_l} with height. 
  }
  \label{fig:typical_length_height}
\end{figure}

Fig.~\ref{fig:typical_length_height} shows the change in $ l_{typ}$ throughout the domain for different metallicity values. Such a typical integral-length-scale is affected by the metallicity value of the simulation at the optical surface as well as in the convective region (see Fig.~\ref{fig:typical_length_height}). With increasing metallicity $ l_{typ}$ increases around the optical surface, which can be associated with larger granules. In addition, metallicity values that are greater and lower than the solar metallicity value lead to slightly greater convection cells.  This result is in agreement with previous studies of pure hydrodynamic convection with different metallicity \citep{Magic_2013A&A}.

\section{Kinetic and magnetic energies}
\label{app:energies}

\begin{figure}
  \centering 
    {\includegraphics[width=1.0\linewidth]{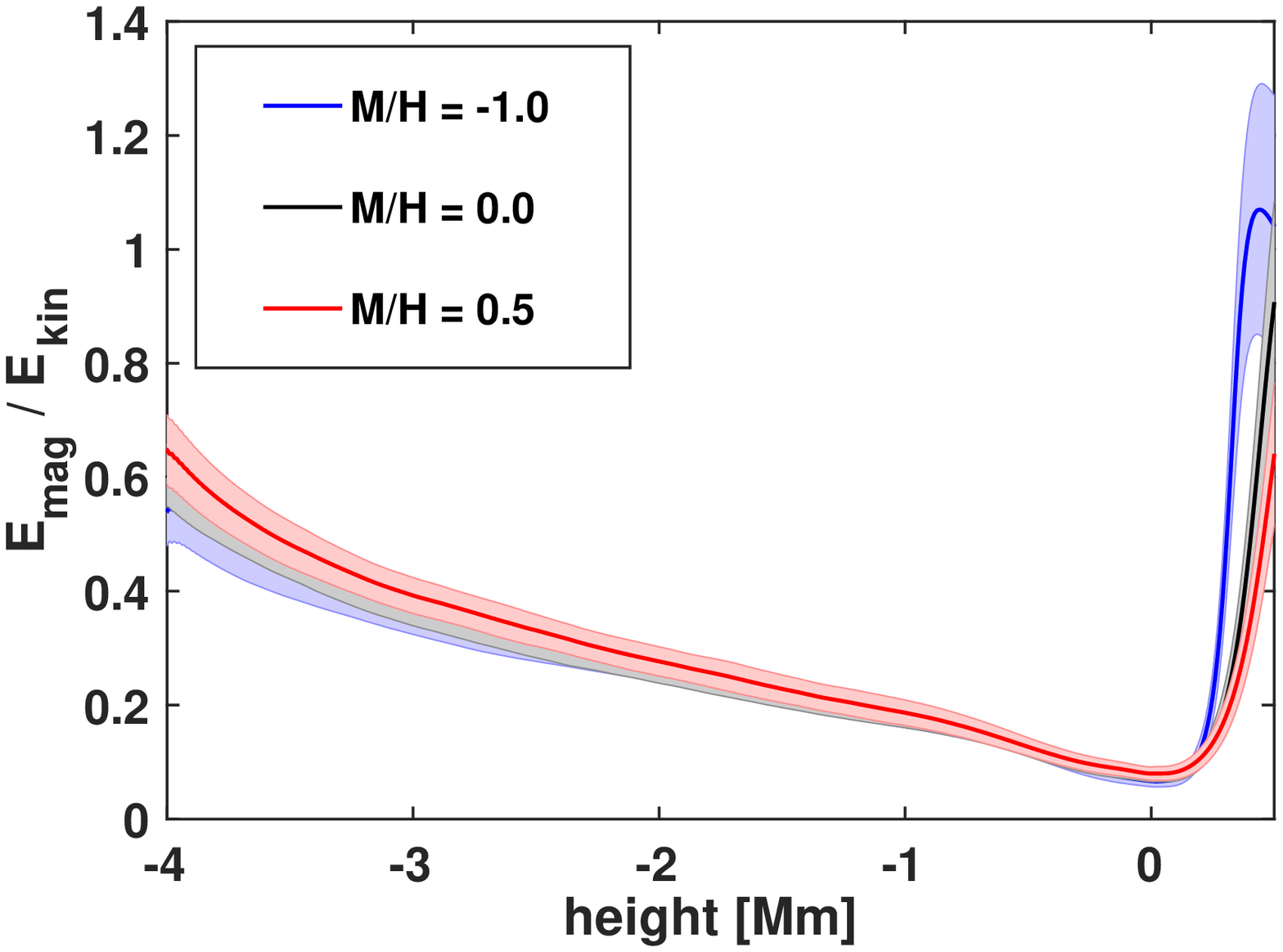}}
  \caption{Ratio of magnetic energy to kinetic energy in the SSD simulations. }
  \label{afig:ratio_emag_ekin}
\end{figure}

 The ratio of magnetic to kinetic energy, shown in Fig.~\ref{afig:ratio_emag_ekin}, suggest that in the deep layers almost an equipartition of the energies is reached. There is a noticeable difference with metallicity, where the ratio is greater for higher metallicity values. While such a ratio can inform about the current state of the SSD simulation, it has no information about how much kinetic energy is potentially converted into magnetic energy.
 
\begin{figure}
  \centering 
    {\includegraphics[width=1.0\linewidth]{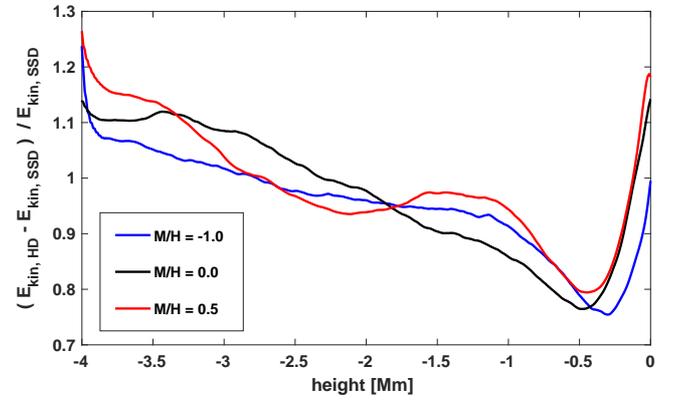}}
  \caption{Ratio of magnetic energy to kinetic energy change from hydro to SSD. }
  \label{afig:efficiency_magn}
\end{figure}

 Therefore, to understand the efficiency of the SSD we calculate the ratio of the change in kinetic energy from the hydro to the SSD run ($\Delta E_{kin} = E_{kin, HD} - E_{kin, SSD}$) compared to the amount of magnetic energy in the SSD simulation. 
Fig.~\ref{afig:efficiency_magn} shows  that overall this ratio is slightly above unity at the bottom of our simulation and drops below unity roughly above 2 Mm. In the upper convection zone up to 2 Mm below the optical surface the magnetic field generation or maintenance is most efficient for M/H = 0.5, while the solar metallicity is the least efficient.  However,  due to the complex behaviour with depth, we conclude that the generation of small-scale magnetic fields is similarly efficient for all three metallicities. The significantly stronger magnetic fields for lower metallicities are mostly due to the higher kinetic energy in these cases.

\end{appendix}
\end{document}